\documentstyle[amssymb,aps,prb]{revtex}
\begin{document}
\input epsf.sty
\twocolumn[\hsize\textwidth\columnwidth\hsize\csname %
@twocolumnfalse\endcsname
\draft
\widetext


\title{First-order nature of the ferromagnetic phase transition in (La-Ca)MnO%
$_{3}$ near optimal doping}
\author{C. P. Adams,$^{1,2}$ \cite{stfx} J. W. Lynn,$^{1,2}$ V. N. Smolyaninova,%
$^{2}$ \cite{towson} A. Biswas,$^{2}$ R. L. Greene,$^{2}$ W. Ratcliff II,$%
^{3} $ S-W. Cheong,$^{3}$ Y. M. Mukovskii,$^{4}$ and D. A. Shulyatev$^{4}$ }
\address{$^{1}$NIST Center for Neutron Research, National Institute of\\
Standards and Technology, Gaithersburg, MD 20899-8562}
\address{$^{2}$Center for Superconductivity Research, Department of Physics,%
\\
University of Maryland, College Park, MD 20742}
\address{$^{3}$Department of Physics and Astronomy, Rutgers University,\\
Piscataway, NJ 08854}
\address{$^{4}$Moscow Steel and Alloys Institute, Moscow 117936, Russia}
\date{\today }
\maketitle

\begin{abstract}
Neutron scattering has been used to study the nature of the ferromagnetic
transition in single crystals of La$_{0.7}$Ca$_{0.3}$MnO$_{3}$ and La$_{0.8}$%
Ca$_{0.2}$MnO$_{3}$, and polycrystalline samples of La$_{0.67}$Ca$_{0.33}$MnO$%
_{3}$ and La$_{5/8}$Ca$_{3/8}$MnO$_{3}$ where the naturally occurring $^{16}$%
O can be replaced with the $^{18}$O isotope. Small angle neutron scattering
on the $x=0.3$ single crystal reveals a discontinuous change in the scattering
at the Curie temperature for wave vectors below $\approx $0.065 \AA $^{-1}$.
Strong relaxation effects are observed for this domain scattering, for the
magnetic order parameter, and for the quasielastic scattering, demonstrating
that the transition is not continuous in nature, in good agreement with the
temperature dependence of the central component of the magnetic fluctuation
spectrum, the polaron correlations, and the spin stiffness reported
previously. This behavior contrasts with the continuous behavior observed
for the $x=0.2$ crystal, which is well away from optimal doping. \ There is a
large oxygen isotope effect observed for the $T_{C}$ in the polycrystalline
samples, and the Curie temperature is decreased by 7 K by substituting 50\%
$^{18}%
{\rm O}$ in the $x=0.33$ sample. \ For the optimally doped $x=3/8$ sample we
observed $T_{C}(^{16}{\rm O})=266.5$~K and $T_{C}(^{18}{\rm O})=261.5$~K
at 90\% $^{18}{\rm O}$ substitution.
Although $T_{C}$ is decreased by 5 K for the $x=3/8$ sample the temperature
dependence of the spin-wave stiffness is found to be identical{\em \ }for
the two samples. These results indicate that $T_{C}$ is not solely
determined by the magnetic subsystem, but instead the ferromagnetic phase is
truncated by the formation of polarons which cause an abrupt transition to
the paramagnetic, insulating state. \ Application of uniaxial stress in the
$x=0.3$ single crystal sharply enhances the polaron scattering at room
temperature. Measurements of the phonon density-of-states show only modest
differences above and below $T_{C}$ and between the two different isotopic
samples.
\end{abstract}


\phantom{.}
]
\narrowtext

\section{Introduction}

One of the simplifying features of conventional isotropic ferromagnets such
as Fe, Co, Ni, and EuO is the isolation of the spin system from the
underlying lattice. This made these types of materials ideal candidates to
investigate the critical dynamics, scaling behavior, and nature of
second-order phase transitions. This situation contrasts dramatically with
the colossal magnetoresistive (CMR) oxides, exemplified by La$_{1-x}$Ca$_{x}$%
MnO$_{3}$, where the electronic, lattice, and magnetic degrees of freedom
are intimately intertwined.\cite{Wollan,Goodenough,tokurascience,%
ramirez97,lynn00,dagottoreview,cheongreview,Theory}
This coupling leads to a number of interesting physical phenomena,
including orbital ordering, charge ordering, polarons, stripes,
ferromagnetic droplet formation at small\cite{Henniondrop,Granadodrop} and
large $x$, and the dramatic decrease in the electrical resistance in the
ferromagnetic-metallic state that is the CMR effect. Investigations of these
phenomena and the close relationship between CMR compounds, relaxor
ferroelectrics, and high temperature superconductors has made this a field
of intense study. An additional strong motivation comes from the possibility
of technical applications of CMR materials in a new generation of magnetic
sensors, read/write heads, and for a variety of new spin-sensitive
electronics.

In the CMR materials the transport and magnetic order are connected through
the ferromagnetic double exchange mechanism\cite{zener51} while the
Jahn-Teller effect locally distorts the lattice that surrounds a Mn$^{3+}$
ion,\cite{Wollan,Goodenough,Theory,zhao96,mori98,billinge96,alexandrov95}
coupling the electronic and magnetic degrees of freedom with the lattice.
The resulting ferromagnetic transition was found to be dramatically
different from conventional isotropic ferromagnets, and was interpreted as a
discontinuous transition between the ferromagnetic-metallic state and the
paramagnetic-insulting state.\cite{lynn96} In particular, the spin wave
stiffness was found not to renormalize to zero at the Curie temperature as
it should in a second-order transition. \ Instead, the spin wave intensities
decreased rather than increasing, and the spectral weight of the magnetic
fluctuations shifted to a quasielastic spin-diffusive peak in the
fluctuation spectrum that rapidly developed as $T\rightarrow T_{C}$ and
peaked near $T_{C}$.\cite{lynn96,fernandez98,adams2000} Concomitant with
this spin-diffusive component is the development of diffuse lattice
(polaron) scattering, that is correlated in a way that is consistent with
the formation of zig-zag, CE-type structural stripes. The intensity for
these polaron-polaron correlations also increases dramatically as $%
T\rightarrow T_{C}$ in a manner very similar to the spin-diffusion
component, and the temperature dependence for both of these closely follows
the resistivity of the metal-insulator transition.\cite%
{lynn00,adams2000,dai2000,doloc99,shimomura99,Kim2000,%
lynn01,Nelson,Kiryukhin2002} It has recently been observed that 
these polarons form a glass above the
ferromagnetic-metallic to paramagnetic-insulator transition, which
subsequently melts at a higher transition temperature to a fluid.\cite%
{Argyriouglass,Lynnglass}

Considering the broad range of compositions, dopings, and crystal structures
that comprise the CMR class of materials, it is not simply the observation
of polarons and charge ordering that is significant, but the establishment
of a clear relationship between the metal-insulator transition and polaron
formation that is crucial. We have continued our neutron scattering
investigations using single crystal samples of La$_{0.7}$Ca$_{0.3}$MnO$_{3}$
and La$_{0.8}$Ca$_{0.2}$MnO$_{3}$, and polycrystalline samples of La$_{0.67}$%
Ca$_{0.33}$MnO$_{3}$ and La$_{5/8}$Ca$_{3/8}$MnO$_{3}$ that have been treated
with $^{16}$O and $^{18}$O isotopes. The results demonstrate a ferromagnetic
transition that is clearly first order in nature, as interpreted in the
original measurements on this system.\cite{lynn96} The results also make it
clear that the observed $T_{C}$ is determined not by magnetic interactions
but instead by polaron formation; the formation of polarons truncates the
ferromagnetic-metallic state. We also show that the application of uniaxial
stress enhances the polaron scattering. \ Finally, we have measured the
phonon density-of-states both above and below $T_{C}$ for both isotopes of
oxygen, and observe small but measurable differences in the lattice dynamics.

\section{Experimental Details}

The single crystal samples of 
La$_{0.7}$Ca$_{0.3}$MnO$_{3}$ and La$_{0.8}$Ca$%
_{0.2}$MnO$_{3}$ were grown by the floating zone technique and have a mass
of $\approx 0.7{\rm \ }$g. The crystal mosaics were single peaked with an
intrinsic width less than a $\frac{1}{4}$ degree. In this range of
compositions the crystal structure is orthorhombic but the distortion is
small and the crystallographic domains appear to be equally populated. \
Therefore, on average, the sample itself is cubic, and we use cubic notation
for simplicity. Nearest-neighbor manganese atoms are then along the
[100]-type directions. The crystals were mounted in the $(hk0)$ plane and
the cubic lattice parameter was $a=3.867$~\AA\ at room temperature. Elastic
and inelastic neutron scattering measurements were collected on the BT-2 and
BT-9 thermal triple axis spectrometers at the NIST Center for Neutron 
Research, using pyrolytic graphite
monochromators, analyzers, and filters. Small angle neutron scattering data
were obtained on the NG-7 30 m SANS spectrometer.

The $x=0.3$ single crystal has a Curie temperature of 251 K, which is the
highest $T_{C}$ for any Ca doped crystals. It has not proved possible yet to
grow crystals with higher 
Ca content, and therefore bulk studies at higher $%
x $ are restricted to polycrystalline samples. For the isotope studies a 20
g polycrystalline sample of La$_{5/8}$Ca$_{3/8}$MnO$_{3}$ was specifically
prepared for the measurements, using the standard solid-state reaction
technique. This Ca concentration was chosen because it is at the maximum $%
T_{C}$ and therefore any change in the Curie temperature cannot be caused by
small changes in the overall oxygen content.\cite{schiffer95,cheongmax} At
each stage of the measurements high resolution powder diffraction
measurements were carried out using the BT-1 powder diffractometer to
characterize the powder sample(s), and at each stage the samples were found
to be single phase. To implement the isotope substitution, the original
sample was divided in half and each half was heat treated in parallel with
either $^{16}$O and $^{18}$O, respectively, through an established 
procedure.%
\cite{zhao96} The fraction of isotopic substitution was determined by
measuring the weight change, which established that approximately 90\%
substitution of $^{18}$O was achieved.\ Bulk magnetization measurements were
taken using a commercial SQUID magnetometer,\cite{smolyaninova01} while the
ferromagnetic order parameter was measured by neutron diffraction. \ The
nuclear scattering amplitudes for $^{16}$O and $^{18}$O are
indistinguishable so that the total oxygen content can be determined by
diffraction, but the fraction of substitution cannot be determined by this
technique. After the $^{18}$O inelastic and diffraction measurements were
completed, the sample was converted back to $^{16}$O and the properties were
re-measured to establish that there was no change in the composition or
changes in properties from the initial sample, within experimental error.
Finally, measurements were also carried out on the original $x=0.33$ powder
sample\cite{lynn96,HuangCacryst} for comparison purposes. This sample was
re-heat treated\ at higher temperatures compared to the original preparation
(and a small amount of Ca may have been lost in this process), and then the
isotope substitution was performed. In these initial measurements only 50\%
substitution was achieved due to the limited amount of $^{18}$O available at
the time, but the shift in $T_{C}$ was still larger than at optimal doping.\ 

BT-2 and BT-9 thermal triple-axis spectrometers were employed to determine
the magnetic order parameters and spin dynamics. For doping $x$ between $%
x\approx 0.15$ and $x\approx 0.5$ La$_{1-x}$Ca$_{x}$MnO$_{3}$ compounds are
ferromagnetic at low temperatures.\cite{schiffer95,cheongmax} 
At smaller $x$
there is anisotropy in the dispersion relations and additional excitations,
\cite{Hennion2001} but in the regime of interest here the system behaves to
a very good approximation as an ideal isotropic ferromagnet. In
polycrystalline samples this makes it possible to measure the spin dynamics
around the forward scattering direction using the (000) peak reciprocal
lattice point,\cite{lynn96,lynn95} while of course in single crystals the
dispersion relations and linewidths can be measured around any reciprocal
lattice point.\cite{doloc98} In either case the long wavelength excitations
in a ferromagnet should have quadratic dispersion 
\begin{equation}
E_{SW}=Dq^{2}+\Delta
\end{equation}%
with a temperature-dependent spin wave stiffness $D$ and a possible spin
wave gap $\Delta $ that would arise from anisotropy. The spin wave spectrum
has been found to be gapless within the precision of neutron inelastic
scattering measurements,\cite{lynn00} which indicates that these materials
are to a very good approximation rotationally isotropic.

The phonon density-of-states was measured on the Filter Analyzer Neutron
Spectrometer (FANS) and the time-of-flight Fermi Chopper Spectrometer (FCS).
Uncertainties quoted for all the neutron measurements are statistical and
represent one standard deviation.

\section{Magnetic Properties for x=0.2}

In Fig. 1(a) we show a measurement of the ferromagnetic (110) Bragg peak
in the vicinity of the Curie temperature for the $x=0.2$ single crystal.
Above the Curie temperature only the Bragg peak due to nuclear scattering is
observed, while below $T_C$ the magnetic contribution to the peak 
intensity is
proportional to square of the order parameter (magnetization). This magnetic
intensity is well described by a power law, except in the immediate vicinity
of the Curie point where critical scattering is significant, and these data
were excluded from the fit. The fit gives $T_{C}=181.04\pm 0.14$~K, and an
(effective) exponent $\beta =0.324\pm 0.007$ which is close to the
prediction of the three-dimensional Heisenberg model ($\beta =0.365$).\cite%
{privman91} The errors represent statistical uncertainties only, and no
account was taken of a possible spread of transition temperatures since the
critical properties are not central to the present investigation. This Curie
temperature is in good agreement with the values of 175~K, 
\cite{okuda2000} 178~K,\cite{dai2001} and
185~K\cite{Hennion2001} reported previously for single crystals of the same
nominal composition.

The spin wave spectrum has been measured at small wave vectors, and the spin
wave stiffness coefficient $D$ is shown in Fig. 1(b) as a function of
temperature. Each constant-$q$ scan was least-squares fit to a quadratic
dispersion law convoluted with the instrumental resolution. Correcting for
the spectrometer resolution has the effect of decreasing the observed value
of $D$ by $\sim 5-10\%$, and the low temperature value was determined to be $%
46\pm 2$~meV-\AA $^{2}$ for this single crystal. This is in good agreement
with the values previously reported for this composition.\cite%
{Hennion2001,okuda2000,dai2001} As the Curie temperature is approached 
the spin wave
intensities were found to increase due to the increase in the thermal
population of spin waves, caused by the combined higher temperatures and
lower $D(T)$ values. In Fig. 1(b) we see that $D$ appears to renormalize to
zero at $T_{C}$, as expected for a conventional second-order 
ferromagnetic transition. We%
\begin{figure}
\hspace{-6mm}
\centerline{\epsfxsize=3.0in\epsfbox{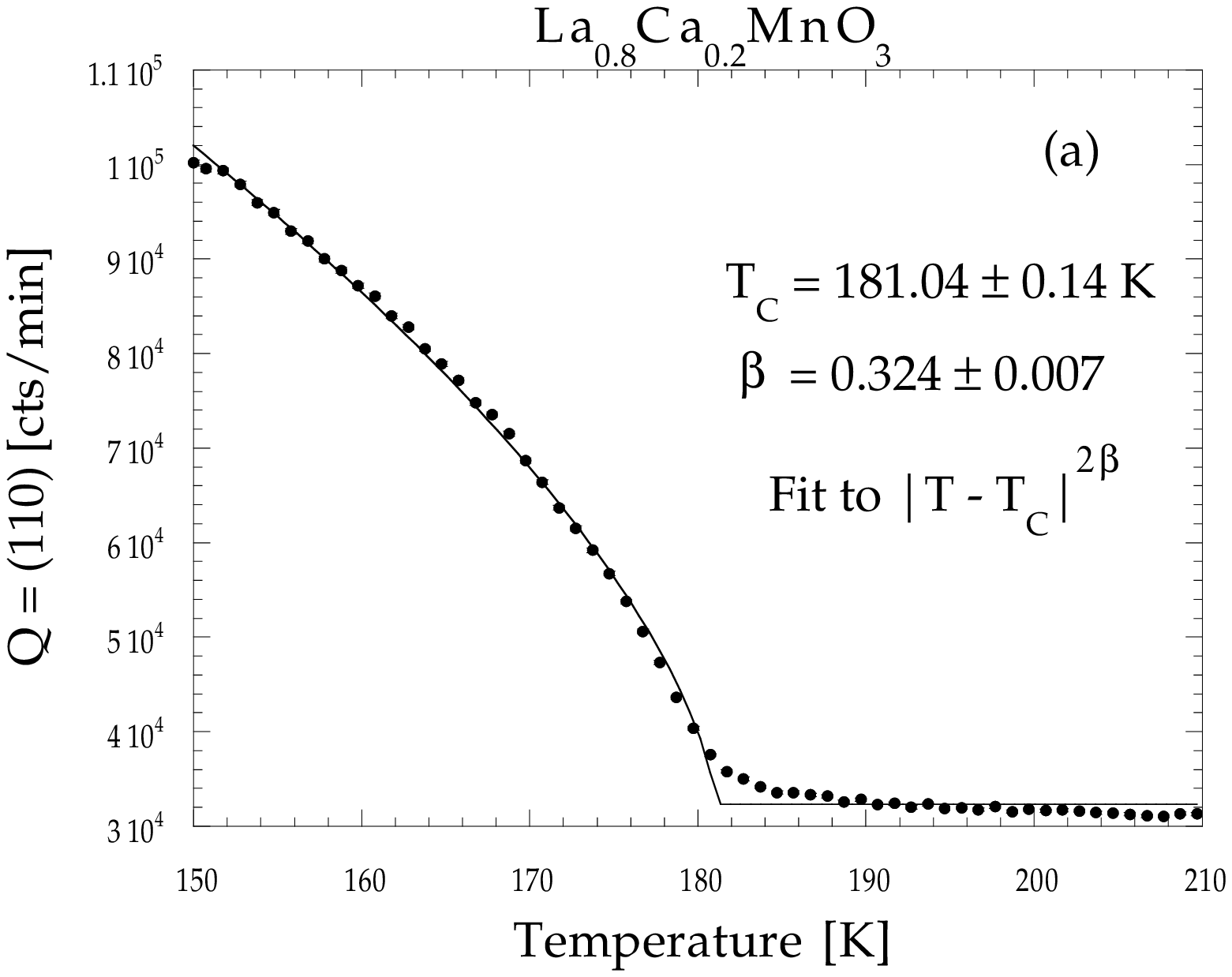}}
\centerline{\epsfxsize=3.1in\epsfbox{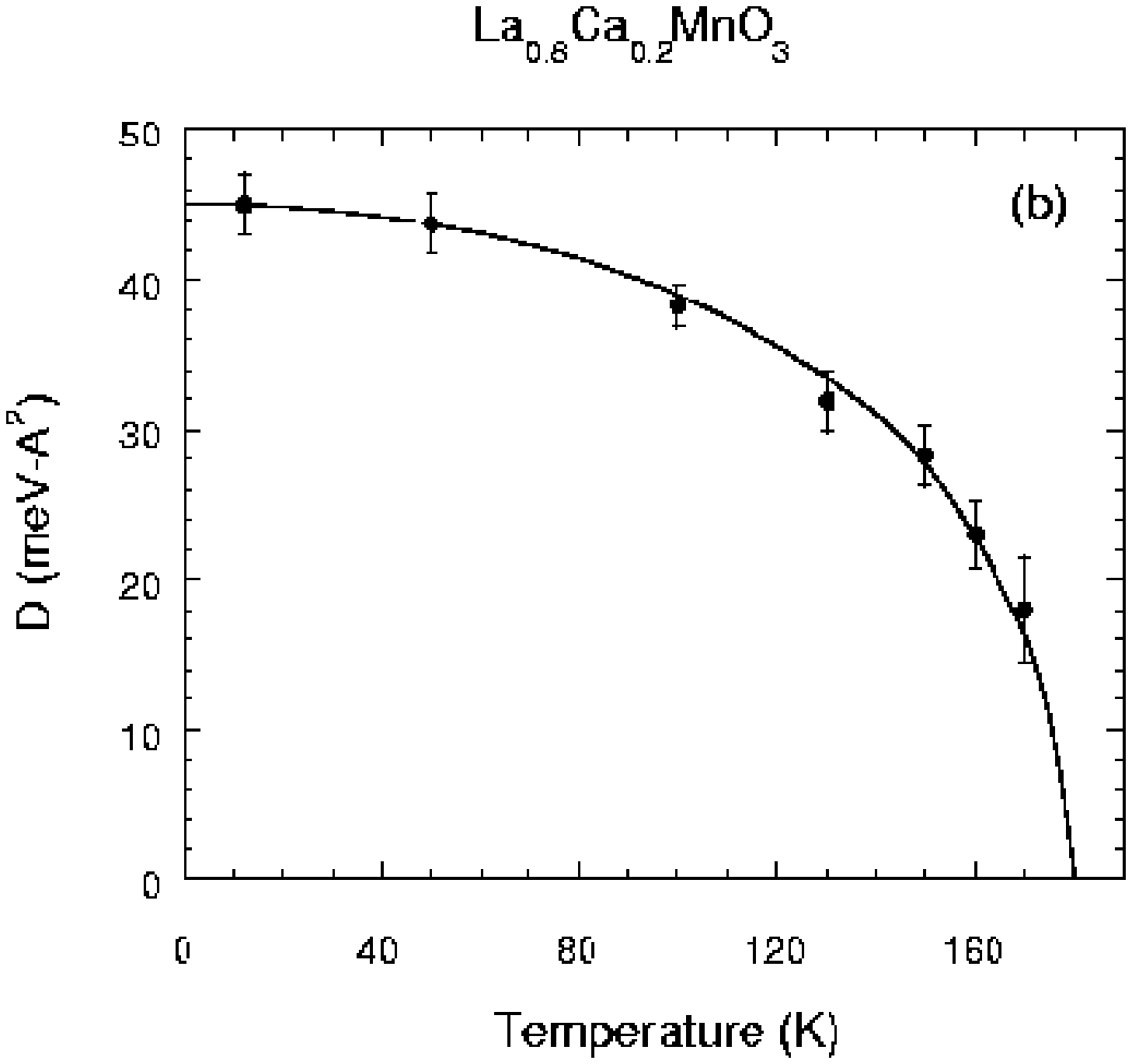}}
\caption[OP and D for Ca 20\%]{(a) Measurement of ferromagnetic order
parameter at the (110) peak for a single crystal of La$_{0.8}$Ca$_{0.2}$MnO%
$_{3}$ near $T_{C}$. Data are well described by a power law with an exponent
that agrees with a three-dimensional Heisenberg model. (b) Spin wave
stiffness for the same sample. $D$ renormalizes to zero as $T\rightarrow
T_{C}$, as expected for a conventional ferromagnetic transition that is
continuous in nature.}
\end{figure}
did not find the development of a central quasielastic
component to the magnetic fluctuation spectrum, in contrast to the behavior
found in the metallic regime.\cite{lynn96,fernandez98,adams2000,dai2001} For
the $x=0.20$ composition, Dai {\em et al. }\cite{dai2001}
also did not find a central
component, but they did suggest that $D(T_{C})$ was finite, which we do not
see in the present measurements.

The overall behavior changes as $x$ increases to the optimal doping range
for ferromagnetism. The excitations at long wavelengths are still isotropic
spin waves with quadratic dispersion, but as mentioned in the introduction $%
D $ fails to renormalize to zero at $T_{C}$. Instead a quasielastic peak
appears in the fluctuation spectrum, indicative of the type of spin
diffusion that occurs above $T_{C}$. This was interpreted as a coexistence
between the low-$T$ ferromagnetic-metallic phase, and the high-$T$
paramagnetic-insulator phase, with the phase fractions changing with
temperature.\cite{lynn96,heffner2times} The spin diffusion component then
grows at the expense of the spin wave intensities. Measurements of the
magnetic correlation length as one approaches $T_{C}$ from above show that
this length does not diverge at the transition as would be expected for a
continuous transition.\cite{lynn96,adams2000,deteresa97} This agrees with
measurements of the paramagnetic scattering, and the correlation length is
only weakly temperature dependent at $\sim $15 ${\rm 
\text{\AA}%
}$ through the transition. These features cannot be explained in the context
of a conventional second-order ferromagnetic phase transition, but can be
understood if one assumes that long range ferromagnetic order is not the
sole order parameter of the transition and that the coupling to the other
order parameter forces the transition to be first order.
\cite{GoodenoughFO96}

\section{Magnetic Properties Near Optimal Doping}

For a conventional ferromagnet above $T_{C}$ the magnetic scattering is
concentrated around the Bragg peaks with $q$-dependent intensity that
follows the Ornstein-Zernike form 
\begin{equation}
I\propto \frac{1}{q^{2}+\kappa ^{2}}\ {\bf .}
\end{equation}%
The correlation length $\xi $ $(=1/\kappa )$ in real space is the parameter
that diverges at $T_{C}$ for a second order transition. Such a divergence
was not observed in previous polycrystalline measurements,\cite%
{lynn96,deteresa97} or in more recent single crystal measurements made
around Bragg peaks.\cite{fernandez98,adams2000} However, SANS measurements on
polycrystalline samples are restricted due to the large amount of small
angle metallurgical scattering from the crystallites, and therefore we have
pursued SANS measurements on the $x=0.3$
\begin{figure}[tbp]
\hspace{-6mm}
\centerline{\epsfxsize=3.0in\epsfbox{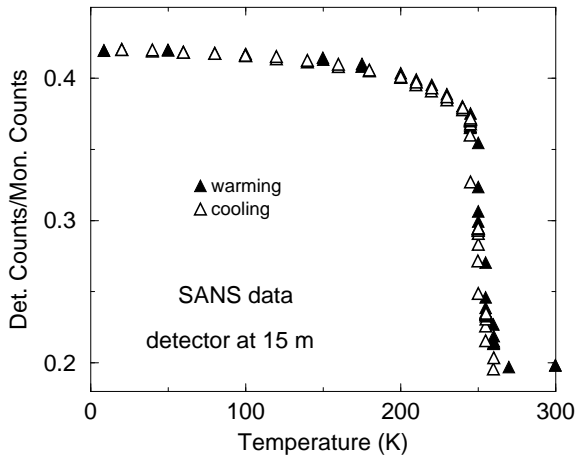}}
\caption[SANS versus T]{Integrated intensity of the small angle neutron
scattering for a single crystal of La$_{0.7}$Ca$_{0.3}$MnO$_{3}$. The
scattering changes in an abrupt manner at $T_{C}=251$~K.}
\end{figure}
single crystal where high quality
data could be obtained over a wide range of wave vectors and temperature,
and where we have also been able to measure the time-dependence of the
scattering.

The results are quite different from conventional second-order behavior, as
(now) expected. The Ornstein-Zernike behavior in Eq. (2) is maintained for $%
q\gtrapprox 0.065$~\AA $^{-1}$ with a short and weakly temperature-dependent
correlation length as found previously.\cite{adams2000} 
For smaller $q$, on
the other hand, there is a discontinuous increase in the scattering at $%
T_{C} $ that cannot be described in terms of the Ornstein-Zernike form or
the common Porod form ($\log I\propto \log q$).\cite{SANSref} This
scattering is strongly peaked in the forward direction, and dominates the
intensity on the (two-dimensional position-sensitive) detector. \ Figure 2
shows the temperature variation of this scattering, recorded as the total
counts on the detector. This corresponds to an integration of the scattered
intensity from roughly $q=0.001$~\AA $^{-1}$ to $q=0.015$~\AA $^{-1}$. The
wavelength is equal to $10\ {\rm 
\text{\AA}%
}$ and the detector was 15 m from the sample for this measurement. The $q$
dependence of the intensity is typical for scattering from domains and
domain walls,\cite{pyrochlore} and the abrupt nature of its onset is
consistent with the transition being discontinuous.

\begin{figure}[tbp]
\hspace{-6mm}
\centerline{\epsfxsize=3.0in\epsfbox{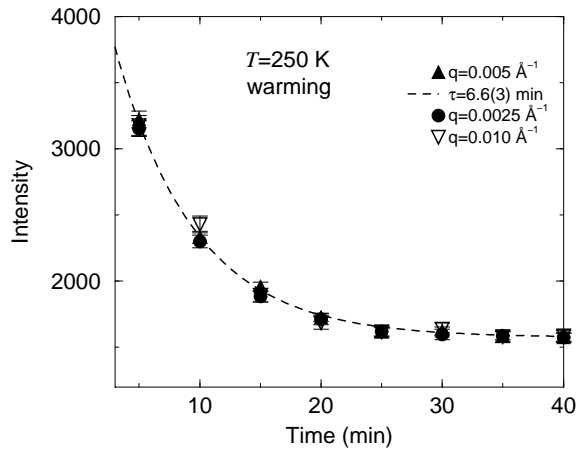}}
\caption[SANS vs t, tau vs T]{Time dependence of the observed SANS
scattering for three different wave vectors at 250 K. Data are scaled at
$t=0$ and shifted vertically to the same asymptotic value, and are fit to a 
decaying exponential with a time constant $\protect%
\tau =6.6\pm 0.3$~minutes. }
\end{figure}
One hallmark of first-order transitions is that, even if thermal hysteresis
is too small to be observed, they are accompanied by relaxation effects.
Figure 3 shows SANS measurements at three different wave vectors as a function
of time, after warming to 250~K. The time for the sample to equilibrate near 
$T_{C}$ is clearly quite long. The data can be described by an exponential
decay with relaxation time $\tau $, and the solid curve is a fit of this
form to the data. Different values of $q$ gave similar values of $\tau $
within statistical uncertainties, showing that all length scales measured in
these SANS experiments exhibit similar relaxation times. Warming was
performed in steps
\begin{figure}[tbp]
\hspace{-6mm}
\centerline{\epsfxsize=3.0in\epsfbox{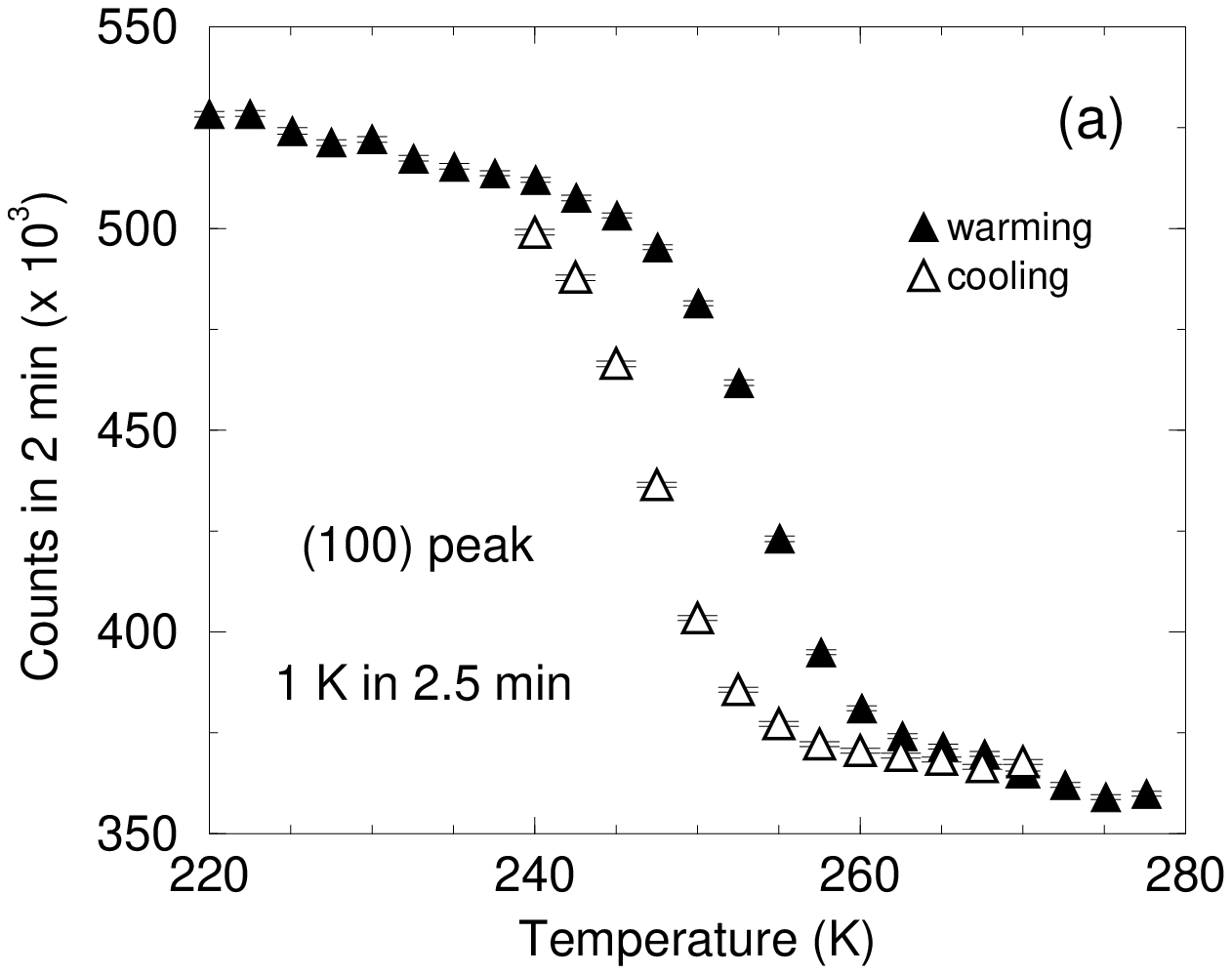}}
\centerline{\epsfxsize=3.0in\epsfbox{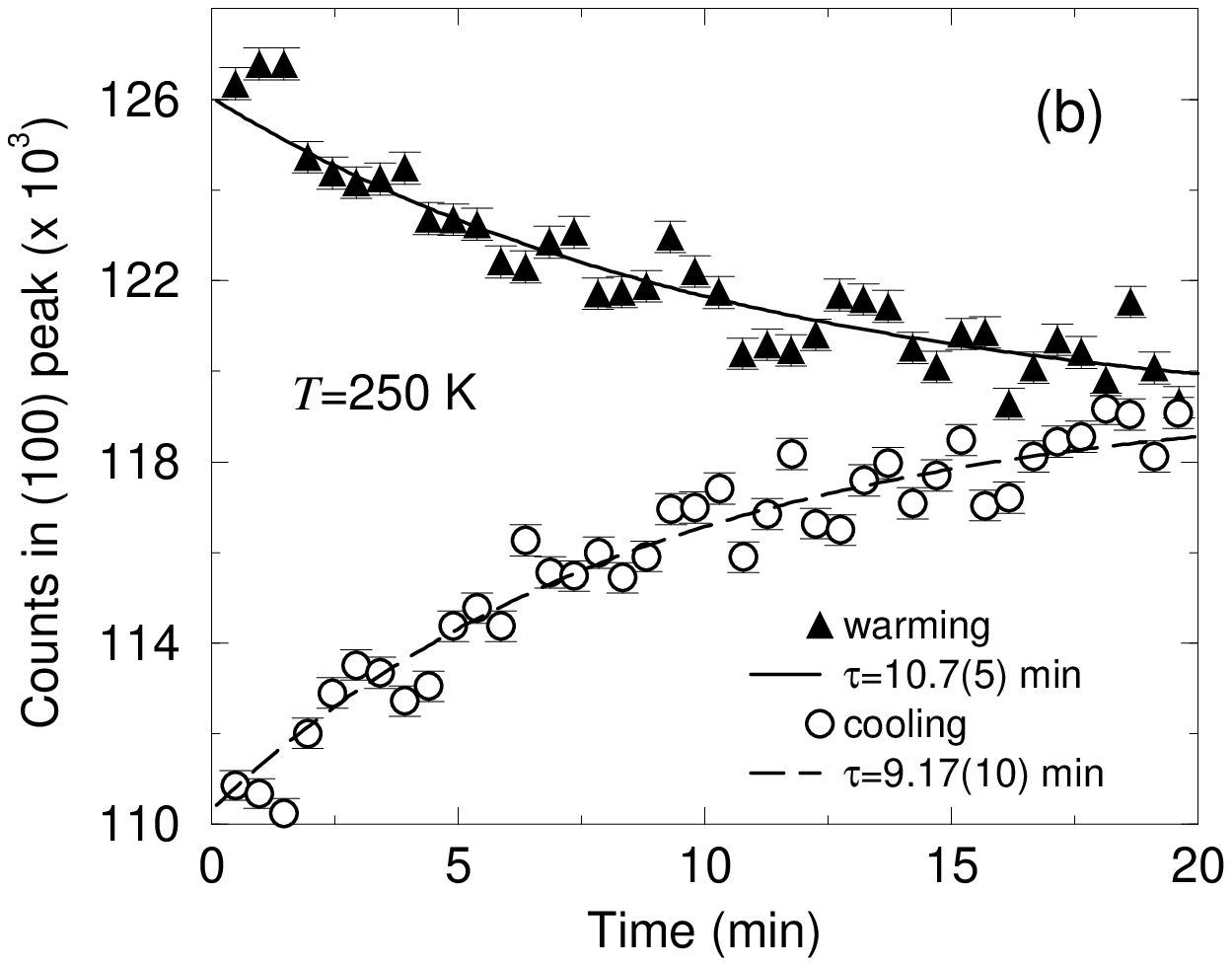}}
\caption[SANS vs t, tau vs T]{(a) \ Thermal irreversibility of the (100)
magnetic Bragg peak on warming and cooling. \ The data were obtained by
changing the temperature in 1 K steps every 2.5 min. (b) Time dependence of
the intensity of the (100) Bragg peak, both warming to the set point of
250 K, and cooling to the same set point. Data are fit to an exponential
relaxation. Note that the time constants are roughly identical and
the asymptotic
values are identical within experimental uncertainties, so there is no
genuine hysteresis observable.}
\label{(100) Bragg peak}
\end{figure}
of 5~K every 40 minutes for these measurements, and the
relaxation time was found to be maximized at $T_{C}$.

One of the difficulties in unambiguously identifying this as a first order
transition has been that, although $D$ does not renormalize to zero, the
magnetic contribution to the intensity of the Bragg peaks appears to be
continuous. \ In particular, although there are some ``irreversibilities''
observed on warming and cooling,\cite{lynn96} no clear hysteresis has been
identified. The availability of the $x=0.30$ single crystal has allowed us
to investigate the behavior of the magnetic Bragg peak in detail. In
Fig. 4(a) we plot the intensity of the (100) Bragg peak, and we see a
clear difference on warming and cooling. This apparent hysteresis is due to
relaxation effects, and in Fig. 4(b) we show the time dependence of the
scattering, after the sample temperature arrives at the set point. \ The
curves are fits to exponential relaxation functions, which describe
the data quite well. \
The fitted $\tau $ versus
\begin{figure}[tbp]
\hspace{-6mm}
\centerline{\epsfxsize=3.0in\epsfbox{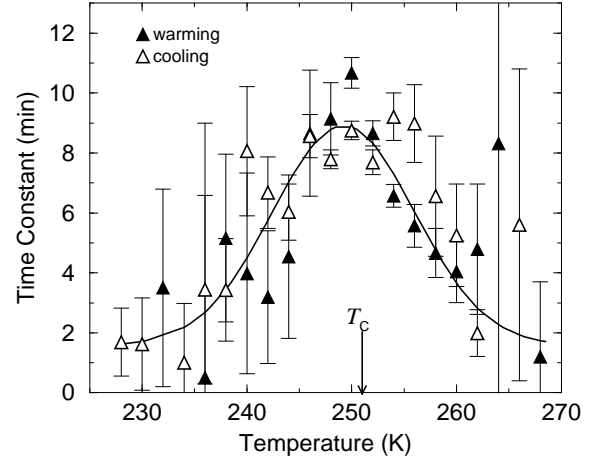}}
\caption[SANS vs t, tau vs T]{Relaxation time $\protect\tau $ as a function
of temperature for the (100) Bragg peak. The short time constant found
away from the transition is typical of the time it takes for a sample to
equilibrate, but in the vicinity of $T_{C}$ the time constant increases
dramatically, indicating that the transition is first order in nature.}
\end{figure}
temperature is shown in Fig. 5, on warming and
cooling. Although the statistical uncertainties are large, particularly when
the total change in scattering between two temperatures is small, we can
clearly see that $\tau $ is maximized near $T_{C}$.

A further result of the time-dependent measurements is that the asymptotic
value of the order parameter at any particular temperature is identical on
warming and cooling, as shown in Fig. 6. \ Thus there is no true thermal
hysteresis in the asymptotic time region. We note that the value for $T_{C}$
determined in these measurements, $251 \pm 3$~K, is slightly lower than the
preliminary value we reported previously $257 \pm 1$~K.\cite{adams2000} The
(correct) value of 251~K is determined here by locating the inflection
point, 
\begin{figure}[tbp]
\hspace{-6mm}
\centerline{\epsfxsize=3.0in\epsfbox{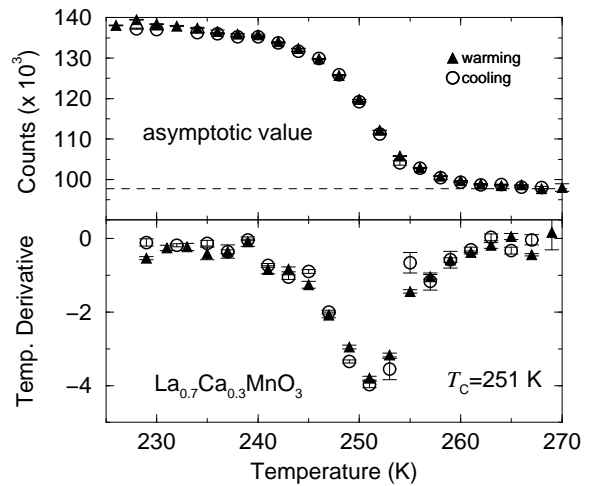}}
\caption[SANS vs t, tau vs T]{Intensity of the (100) magnetic Bragg peak,
in the long-time limit. The data reveal that there is no genuine hysteresis
observable. \ The bottom curve shows the derivative of the data, where the
minimum identifies a Curie temperature of 251 K. }
\end{figure}
\begin{figure}[tbp]
\hspace{-6mm}
\centerline{\epsfxsize=3.0in\epsfbox{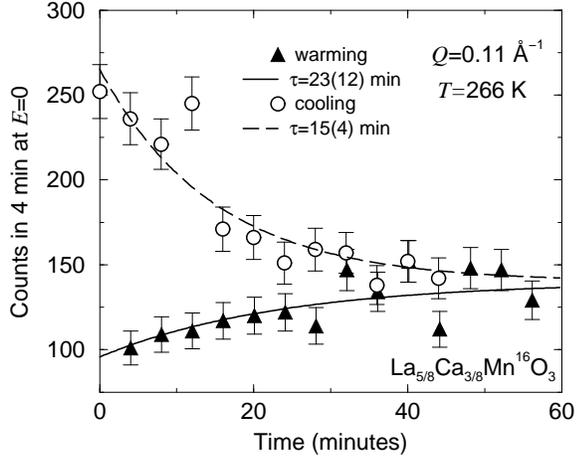}}
\caption[SANS vs t, tau vs T]{Intensity of the quasielastic magnetic
scattering for a temperature of 266 K at a wave vector of $0.11$~\AA $^{-1}.$
\ Note that it takes about an hour for the intensities to become equal on
approaching from low versus high temperatures. \ In the long-time limit
there is no genuine hysteresis observable. }
\label{quasielastic time dependence}
\end{figure}
after taking the data on warming and cooling
and properly taking into
account the long time-scale behavior of the system. This value is also in
good agreement with the abrupt onset of SANS shown in Fig. 2. The SANS data
also lack true thermal hysteresis.

\ We also measured the relaxation of the central quasielastic component of
the magnetic fluctuation spectrum just below $T_{C}$ for the 
$x=3/8$ $^{16}$O sample, and the data are shown in Fig. 7. 
The sample temperature was changed to
266~K from well below or well above this temperature, and the data
collection at the elastic position was started when the sample block reached
the set point. The measured relaxation times are somewhat different for the
two cases, but it is clear that the asymptotic values are identical within
statistical uncertainties. \ Hence, the activation barrier and latent heat
for this transition are rather small, which has hindered the identification
of this as a first order transition in the specific heat until recently.
\cite{Hellman}  Previous measurements of the order parameter (and 
measurements
reported in the next section on polycrystalline samples) had always shown
some irreversibility on warming and cooling as a result of the activation
barrier, but the temperature variation in the order parameter always
appeared to be continuous.

\subsection{Effect of uniaxial stress}

One of the characteristic features of CMR materials is their extraordinary
sensitivity to external perturbations. The application of a magnetic field,
for example, represents a very modest energy in a system that orders at 250
K, but a magnetic field can drive the metal-insulator transition and the
associated CMR effect. Stress, which is always present in thin films, also
has a dramatic effect on the electronic properties, and is one of the
primary reasons for the rejuvenation of interest in this class of materials.
\cite{CMRrefs} We have already established that the formation of polarons in
the paramagnetic state is directly related to the resistivity, 
\cite{adams2000} and we may then 
anticipate that the polarons themselves will be
directly affected by stress.

To date there have been only a few studies of CMR as a function of applied
force, using two-dimensional strain in thin films\cite{2dstrain} and
hydrostatic pressure.\cite{okuda2000,hydro} In addition, the effects 
of internal
pressure that result from substitution of smaller Pr ions for 
the La A-site cation, for example, have been performed. Other 
options for cation
doping have shown that the behavior of the entire $x-T$ La-Ca phase diagram
can be achieved via changing the average radius at the A-site.
\cite{GoodenoughFO96,hwang95,bandwidthCMR} Results can often be 
understood in
terms of the double exchange model when the distortion of the ideal
perovskite lattice is considered. Changing the average radius of the A-site
cation changes the tilt of the oxygen octahedra that surround the B-site
cation, which effects the electronic band width and the double exchange. An
increase in distortion away from the ideal perovskite structure tends to
favor charge-ordered insulating behavior, where the antiferromagnetic
interaction dominates over metallic ferromagnetism.

Our single crystal of La$_{0.7}$Ca$_{0.3}$MnO$_{3}$ was mounted in a stress
rig on the BT-9 triple axis spectrometer in the $(h\/k\/0)$ zone. Stress was
applied along the axis of the crystal, roughly 30 degrees from (001) in
the direction of (220). All measurements were made at room temperature
where the charge peak intensity has dropped to roughly 30\% of its maximum
value near $T_{C}$. Given the geometric limitations imposed by the stress
rig we were essentially limited to measurements near the (220) peak using an
incident energy of 52 meV. The charge order peaks are broad (corresponding
to short range correlations) and occur at wave vectors ($\pm \frac{1}{4}
$, $\pm \frac{1}{4}$, 0). Near (220) these occur at (1.75 2.25 0) and (2.25
1.75 0); the structure factor is negligible for the ``longitudinal'' peaks
at (1.75 1.75 0) and (2.25 2.25 0). In Fig. 8 we show a transverse scan of
the charge order peak at (2.25 1.75 0) at 0 kbar and 1 kbar. The intensity
of the peak has increased by nearly a factor of two, while the intensity of
the fundamental Bragg peaks are unaffected at this level of stress. It is
evident that the diffuse scattering, which is related to the
``single-polaron'' diffuse scattering, has also increased. Uniaxial stress
therefore has a rather dramatic effect on the polaron scattering.

Because of the rather large change in intensity that was observed when 1
kbar of stress was applied, we reduced the force to zero in order to
determine the behavior at lower values of the stress. The polaron intensity
was reduced, but not to the initial value; \ the observed scattering was
about half way between the initial data and the 1 kbar data. \ Subsequent
measurements up to 0.5 kbar showed no change, and with further
increase the scattering began to increase up to the 1 kbar data. Hence we
believe that the scattering would increase approximately linearly with
stress from the virgin state. Unfortunately, 
\begin{figure}[tbp]
\hspace{-6mm}
\centerline{\epsfxsize=3.0in\epsfbox{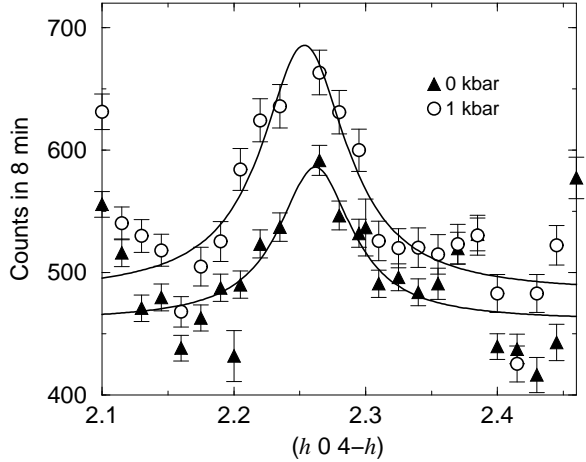}}
\caption[charge peak near (220) at 0 and 1 kbar]{Change in the intensity of
the charge-order peak when uniaxial stress is applied. \ The intensity
increases by more than a factor of two in going from 0 kbar to 1 kbar of
uniaxial stress. The intensity of the single-polaron (diffuse) scattering
also increases with applied stress.}
\end{figure}
shortly after returning to 1
kbar of stress, the crystal shattered, ending the data collection.

\subsection{Oxygen Isotope Effects}

One very strong indication for some type of coupled transition in La$_{1-x}$%
Ca$_{x}$MnO$_{3}$ is the oxygen isotope effect.\cite%
{zhao96,isaac98,Babushkina,franck98,ZhouGood,Heilman,Babushkina2000,Fisher}
Conventionally, magnetism is a property that emerges from the electronic
structure of the system, which is determined by the component elements and
the crystal structure that they assume. Once that crystal structure is
determined, the specific isotopic masses of the components do not usually
enter into the electronic Hamiltonian. The classic example that runs
contrary to this is BCS (Bardeen-Cooper-Schriffer) superconductors where the
phonons lead to an attractive interaction between the electrons and
fundamentally change the nature of the low temperature electronic structure
(a gapped superconductor rather than a Fermi liquid). Such phonon effects
are limited to temperatures less than 40 K. However, in La$_{0.8}$Ca$_{0.2}$%
MnO$_{3}$ substitution of $^{18}$O for $^{16}$O leads to a 20 K decrease in $%
T_{C}$, from 180 K to 160 K.\cite{zhao96}  An isotope effect in
antiferromagnetic La$_{0.5}$Ca$_{0.5}$MnO$_{3}$ is also seen, but in the
opposite direction and effecting the N\'{e}el temperature.\cite{isaac98}

To investigate the effect of oxygen isotope substitution on the magnetic
behavior we chose to work at $x=3/8$, where $T_{C}$ is maximized and the CMR
effect is large.\cite{cheongmax}  This choice of $x$ minimizes any possible
variation in or distribution of $T_{C}$ with overall oxygen content. Indeed,
for the measurements in Ref. \onlinecite{zhao96} 
the observed oxygen isotope
effect has been partially attributed to different oxygen stoichiometry.\cite%
{franck98} For the present samples the high resolution powder diffraction
measurements could be used to establish that the oxygen contents for both
samples
\begin{figure}[tbp]
\hspace{-6mm}
\centerline{\epsfxsize=3.0in\epsfbox{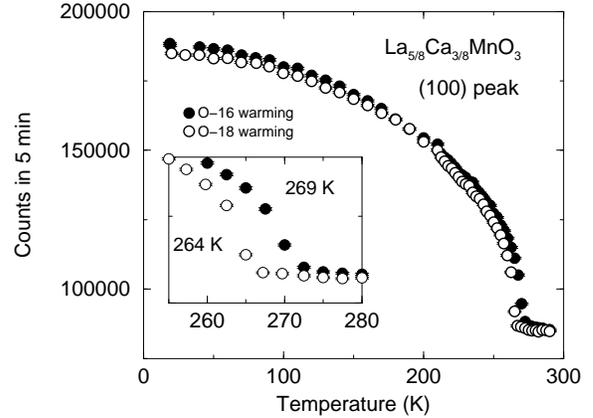}}
\caption[OP for O-16, O-18]{Measurement of the intensity of the magnetic
scattering on warming for polycrystalline samples of La$_{5/8}$Ca$_{3/8}$MnO$%
_{3}$ that have been treated with $^{16}$O and $^{18}$O. The Curie
temperature is reduced by 5~K for the $^{18}$O sample.}
\end{figure}
were identical within experimental uncertainties, and thus there can
be no significant variation in $T_{C}$ due to the total oxygen content. The
lattice parameters, average atomic positions and occupancies were identical
as well. Density measurements indicate that roughly 90\% of the $^{16}$O was
replaced by $^{18}$O in the substituted sample.

Figure 9 shows the intensity of the (100) powder peak as a function of
temperature for the two samples, to determine the effect of oxygen isotope
substitution on $T_{C}$. The data were obtained as a function of warming at
identical heating rates. The Curie temperature of the $^{18}$O sample is 5~K
lower than the $^{16}$O sample, as can been seen in the inset. The same 5~K
difference was seen on cooling, along with a 5~K irreversibility in both the
curves relative to the warming runs. Given the first-order nature of the
transition the irreversibility is expected and is consistent with previous
measurements. Taking the average of these warming and cooling runs gives $%
T_{C}(^{18}{\rm O})=261.5\pm 0.5$~K and $T_{C}(^{16}{\rm O})=266.5\pm 0.5$%
~K. As we pointed out earlier for the single crystal measurements, these $%
T_{C}$'s are determined from the inflection points of the order parameter
curves. The value of $T_{C}$ for the $^{16}$O sample is indeed the maximum
value seen on the Ca-concentration phase diagram and is a further indication
of the high quality of these samples.

This 5 K shift is much smaller than reported in the $x=0.2$ sample, so for
comparison we have also made measurements on the original $x=0.3$ sample.\
With only 50\% $^{18}$O substitution we obtained $\Delta T_{C}=-7$~K. It is
noteworthy that the transition for the partially substituted sample was just
as sharp as for the fully substituted sample, indicating that there is no
significant effect on the properties caused by the oxygen isotopic
randomness. The larger isotopic shift with decreasing $x$ that we find is in
good agreement with the trends already established for this system.

Having confirmed that oxygen isotope substitution changes $T_{C}$, the next
question is how the magnetic interaction strength changes, which we
characterize by
\begin{figure}[tbp]
\hspace{-6mm}
\centerline{\epsfxsize=3.0in\epsfbox{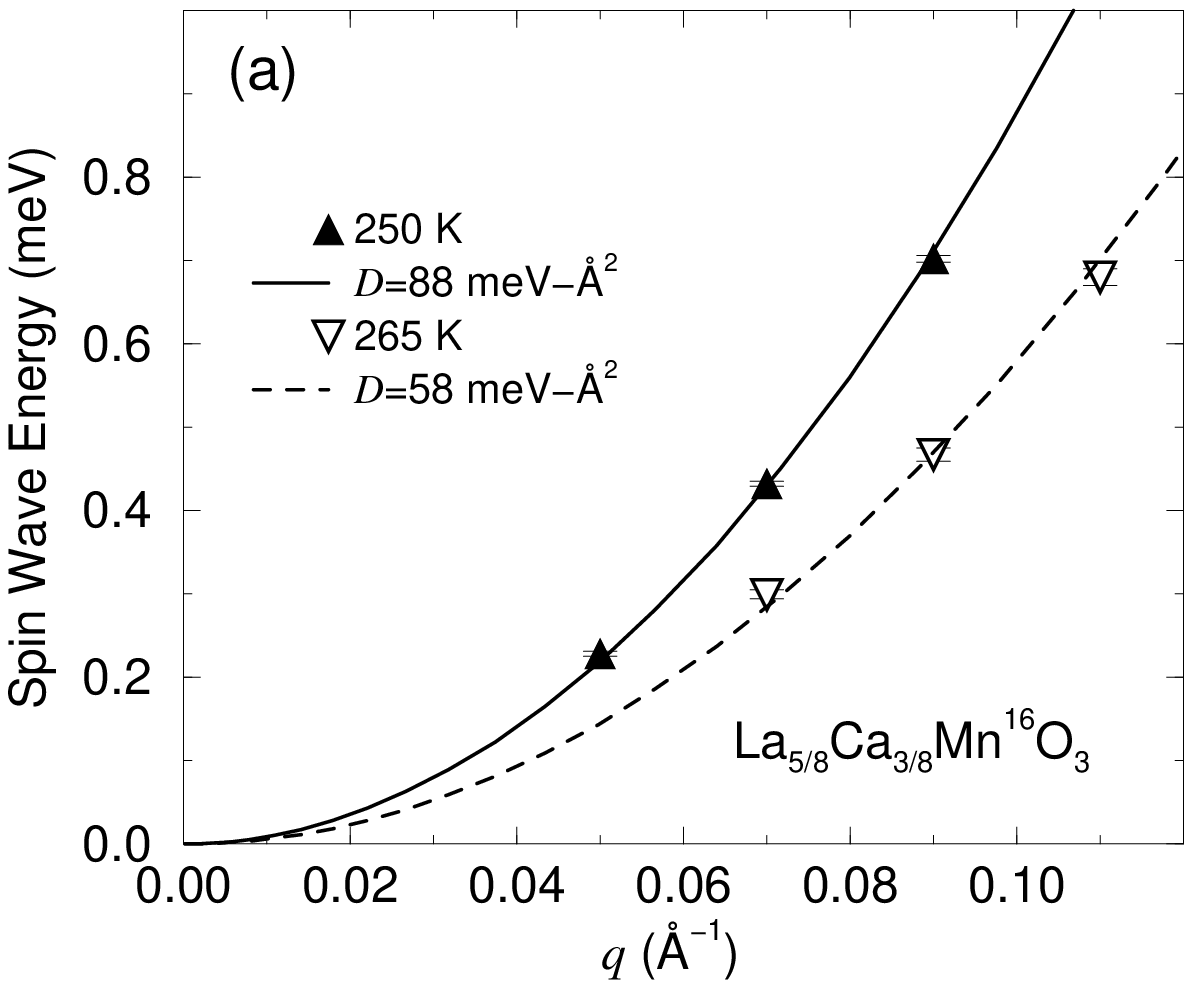}}
\centerline{\epsfxsize=3.0in\epsfbox{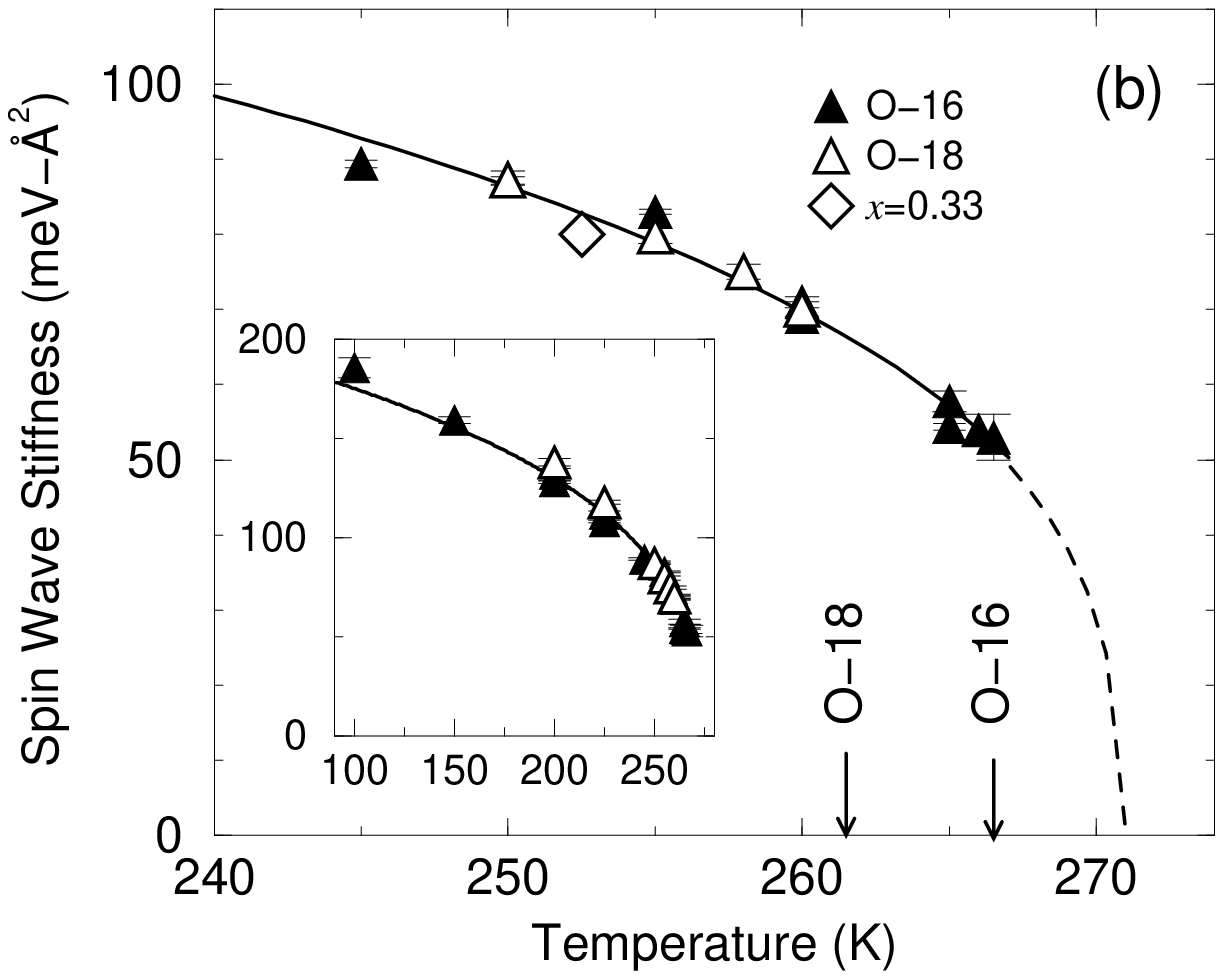}}
\caption[ew vs q, D vs T]{(a) Example of the spin wave dispersion relation
for two typical temperatures. Data are well described by gapless quadratic
dispersion with the indicated values of $D$. (b) $D$ versus temperature for
the $^{16}$O and $^{18}$O samples. The observed $D$ values
are identical at each temperature independent of the oxygen isotope, despite
the difference in $T_{C}$ (as indicated by the arrows). The extrapolated
temperature where $D(T)$ would renormalize to zero is 271 K, but the
ferromagnetic-metallic state abruptly disappears at the temperature where
the polarons form, and this truncation occurs at a lower temperature for the 
$^{18}$O sample. The $D(T_{C})$ value for the original data in Ref.
\onlinecite{lynn96} with $x=0.33$ is also indicated by the open
diamond symbol. 
The inset shows the
stiffness over a wider temperature range. }
\label{truncation}
\end{figure}
measuring the spin dynamics. In Fig. 10(a) we show the low-$%
q $ spin wave dispersion in the $^{16}$O sample at 250~K and 265~K. The
dispersion is well described by Eq. (1) with gapless excitations,
typical of these isotropic ferromagnets.\cite%
{lynn96,fernandez98,adams2000,lynn01,rhyne} The spin wave stiffness has
decreased considerably over this temperature range upon approaching $T_{C}$.
Similar measurements were made to determine $D$ in the $^{18}$O sample. 
Figure
10(b) shows a comparison of $D$ versus $T$ for the two samples. The two
samples show {\em identical} results for $D$ over the entire temperature
range where spin waves are observed. Note also that there are in fact two
complete sets of data for the $^{16}$O, one taken with the two samples heat
treated in parallel, and one set taken after exchanging the $^{18}$O for $%
^{16}$O. Both sets of data are identical. The only difference found is that
for the lower-$T_{C}$ $^{18}$O sample the measurements cannot be continued
to as high a temperature as with the $^{16}$O sample. This result obviously
cannot be understood in the conventional context of ferromagnetism since $%
T_{C}$ has changed but the ferromagnetic interaction as measured by $D$ has
not. Hence it is clear that $T_{C}$ is not determined by magnetic
interactions alone.

The $D(T)$ data can be well described by a power law in this temperature
range, with an extrapolated value of $T_{C}=271$~K which is identical for
both isotopes. The exponent used in this fit is $\beta -\nu =0.4$, which is
fairly close to the prediction of the Heisenberg model $\beta -\nu =0.324$.%
\cite{privman91} The particular value of the exponent is not the central
issue here; the point is that the exchange interaction is not
affected by the isotope substitution, and the extrapolated $T_{C}$ is
identical for both samples. For both samples the actual magnetic transition
occurs before $D$ has renormalized to zero, but note that the
renormalization has progressed much farther than is the case for samples
with lower $x$, lower $T_{C}$ samples. Indeed, Fernandez-Baca {\em et al.}%
\cite{fernandez98} have established that $D$ changes little over a wide
range of tolerance factor, while large changes in $T_{C}$ occur. From this
point of view the (O isotope-independent) magnetic interactions determine
this extrapolated $T_{C}$, but polaron formation truncates the transition,
thereby determining the observed $T_{C}$. This truncation occurs in a
first-order manner and leaves $D$ at a finite value, and is the determining
factor for $T_{C}$ over a range of $x$. Other measurements of $D(T)$ for
smaller $x\sim 0.3$ compounds are very similar to the $x=3/8$ samples, just
with a different truncation value at the different $T_{C}$'s. The downturn
in $T_{C}$ with decreasing $x$ in the phase diagram is then caused by
polaron formation rather than a weakening of the double exchange
mechanism. This behavior continues until the system becomes insulating, and $%
D(T=0 \: {\rm K})$ suddenly drops. For our insulating $x=0.2$ 
sample $D(T=0)$ is 3
times smaller than in the optimally doped sample, in good agreement with the
results of Refs. \onlinecite{dai2001} and \onlinecite{fernandez2002}.

This rapid truncation of ferromagnetism is clearly revealed in Fig. 11,
which shows the magnetic fluctuation spectrum as a function of temperature
at a wave vector $q=0.11$~\AA $^{-1}$ for the two samples. The elastic
incoherent nuclear scattering and a flat temperature independent background
have been removed from these spectra, and the solid curves are fits to the
spin wave dispersion [Eq. (1)] plus a quasielastic (spin diffusion) spectral
weight function convoluted with the instrumental resolution. Spectra at a
series of $q$'s were obtained at each temperature, and the data collection
time between runs at different temperatures is sufficiently long that there
are no irreversibility effects due to warming and cooling. For the $^{16}$O
sample [Fig. 11(a)] we see a three component spectrum at 265~K, just below $%
T_{C}$. The spin waves are observed in energy gain 
($E$%
\mbox{$<$}%
0) and energy loss ($E$%
\mbox{$>$}%
0)
\begin{figure}[tbp]
\hspace{-6mm}
\centerline{\epsfxsize=3.0in\epsfbox{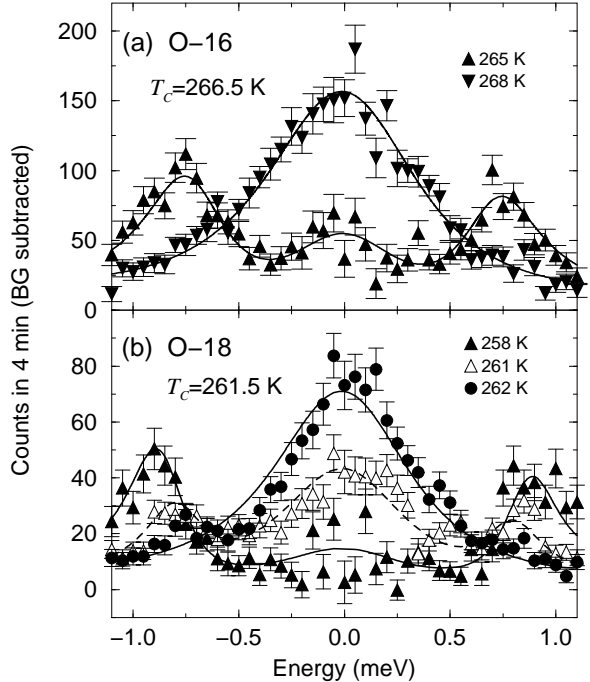}}
\caption[const-q at diff T, x=3/8]{Constant-$q$ scans at a wave vector of $%
0.11$~\AA $^{-1}$ showing the nature of the magnetic fluctuation spectrum in
the vicinity of $T_{C}$ for the $^{16}$O and $^{18}$O samples. (a) For the 
$^{16}$O sample there is a three peak structure at 265 K, while just above $%
T_{C}$ at 268 there is a single quasielastic peak. \ (b) For the $^{18}$O
sample at 258 K most of the weight of the magnetic scattering is in the spin
wave excitations. \ At 261 K the weight in the spin wave scattering has
decreased, and the spectral weight has shifted to the central component. \
At 262 K the scattering is completely diffusive in nature. \ There is no
evidence of the usual collapse of the spin wave energies found in a second
order transition. \ Note that the count rates for the two samples 
are different even 
though the samples were approximately the same size, and this difference is 
due to the data being collected under different experimental 
conditions; $^{16}$O data were taken on BT-2 with collimation of 
20'-10'-10'-20', while the $^{18}$O data were collected on BT-9 with 
collimation of 10'-13'-10'-16'.}
\label{constantqO16O18}
\end{figure}
at $\sim $0.8~meV. At 268~K, just above $T_{C}$, the spin waves have
disappeared and the scattering is purely quasielastic (peaked at $E=0$). For
the $^{18}$O sample [Fig. 11(b)], on the other hand, we observe well defined
spin waves below $T_{C}$ at 258~K along with a quasielastic component fitted
to the data. At 261~K the central component has increased to form a clear
peak (at this $q$), while the spin waves have lost over 50\% of their
intensity and have moved to lower energies, consistent with all previous
observations of the lower $x$ compounds which show growth of the central
peak at the expense of the spin waves. But further increasing temperature by
only 1~K wipes out the spin waves entirely, leaving only a central peak that
corresponds to ferromagnetic correlations in the paramagnetic phase. This
very rapid temperature evolution is evidence for the truncation of the
magnetically ordered phase in a first-order manner. The nature of the
magnetic fluctuation spectrum is the same for both samples, except that
the transition for the $^{16}$O sample is extended higher in 
temperature by 5~K,
which results in a $D(T_{C})$ value that is about 30\% lower than for the $%
^{18}$O sample. Note in particular that the data for the $^{18}$O sample is
pure spin diffusion for 262~K, while for the $^{16}$O sample we have
propagating excitations at 265~K. We remark that we have taken data for
temperatures between the ones shown in Fig. 11 to examine the possibility
that there is a sudden decrease in the spin wave energies that might be
attributable to a sudden decrease in the effective exchange energy, but all
the data indicate that we have two-phase coexistence,\cite{lynn96} a
ferromagnetic-metallic phase and a paramagnetic-polaron glass insulator. For
lower $x$ samples the truncation occurs at lower temperature, with a
corresponding higher value of $D(T_{C})$ as indicated in Fig. 10(b). Thus
the transition at lower $x$ is more strongly first order. We have also found
that the central peak usually was observable over a wider range of
temperature at smaller $x$, 15~K or more, compared to the $\sim $5~K range
observed here. This may also be related to how close the system is to the
extrapolated second-order phase transition temperature $T_{C}$.\cite%
{Hellman,rhyne}

\subsection{Phonon Density-of-States}

With the strong evidence that polaron formation plays a crucial role in the
ferromagnetic-metallic/paramagnetic-insulator phase transition, the question
becomes what controls polaron formation. Given the general relationship
between isotope effects and lattice dynamics, and the electron-phonon
coupling through the Jahn-Teller effect,\cite{millis96} we decided to
measure the phonon density-of-states in our samples using time-of-flight and
filter analyzer spectrometer techniques.

The time-of-flight Fermi chopper spectrometer uses cold monochromatic
neutrons incident on the sample, and then detects the inelastically
scattered neutrons in a bank of detectors. Near room temperature one can
reliably measure energy transfers up to $\sim 30$~meV. The density-of-states
is obtained by removing the thermal population factor, weighting the
scattering by $q^{2}$ (assumes phonon scattering), and then averaging over a
range of $q$. Note that for low energies and modest values of $q$ a
considerable part of the spectrum is magnetic in origin for
these samples.
We have made no attempt to separate out the magnetic contribution from the
lattice (nuclear) contribution.

The filter analyzer spectrometer (FANS) uses cooled polycrystalline
beryllium and polycrystalline graphite as an analyzer, so that only neutrons
with energies below 1.7 meV reach the detectors. The detectors span a large
solid angle (1.3 steradians), and the incident energy is scanned to
determine the inelastic scattering. In contrast to the Fermi chopper
spectrometer this instrument is designed to operate in energy loss mode and
no correction needs
\begin{figure}[tbp]
\hspace{-6mm}
\centerline{\epsfxsize=3.0in\epsfbox{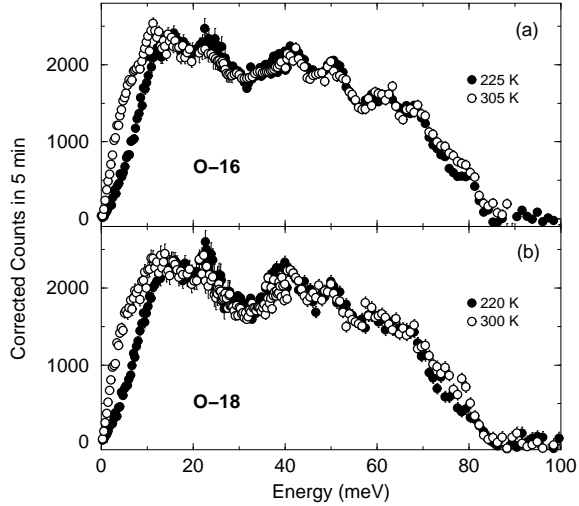}}
\caption[DOS for O-16 and O-18, T-comp]{Phonon density-of-states for (a) $%
^{16}$O and (b) $^{18}$O samples taken above and below $T_{C}$. Units are
those appropriate for the FANS spectrometer with a Cu(220) monochromator.}
\end{figure}
to be is made for $q^{2}$ since a $1/q^{2}$ term enters
into the resolution volume. A run with an empty sample can was used to
remove background. We used a pyrolytic graphite monochromator over the range
of 15 meV to 40 meV and a Cu(220) monochromator for energies between 38 meV
and 120 meV. Data sets for each sample were merged together after background
subtraction. A similar merging was performed to include the low-energy
time-of-flight data. After merging, the data were corrected for multiphonon
effects, which tended to be significant only above 75 meV. Additional
experimental details are given in Ref. \onlinecite{Sumarlin}.

Results for the phonon density-of-states taken above and below $T_{C}$
appear in Fig. 12. Identifiable peaks appear in the data at 17, 25, 42, 51,
and 63 meV. 
There are no phonons above 85 meV after multiphonon corrections. These
observations are in agreement with recent single-crystal measurements of the
phonon spectra of La$_{0.7}$Ca$_{0.3}$MnO$_{3}$, which show no phonons above
72 meV and substantial flat regions in the dispersion branches at 23 meV and
50 meV and an optical band near 40 meV.\cite{ZhangPhonon} Corresponding
peaks in Raman scattering for an $x=0.33$ sample are found at %
%
17, 25, and 51 meV.\cite{granado01} Other Raman peaks are also observed at
8, 10, 30, and 54 meV with a broad maximum at 85 meV. Single crystal phonon
dispersion measurements on the related CMR compound La$_{0.7}$Sr$_{0.3}$MnO$%
_{3}$ show phonon branches up to 72 meV with no gaps between branches when
averaging over the entire zone.\cite{reichardt99} Substantial magnon-phonon
interactions have also been reported,\cite{dai2000prb} but it is likely that
these effects can only be observed in single crystals. For the present
density-of-states data at low energies there are substantial differences
above and below $T_{C}$, but this originates from the change in the magnetic
scattering from spin diffusion above $T_{C}$ to propagating spin waves
below. At higher energies the biggest difference is in the
\begin{figure}[tbp]
\hspace{-6mm}
\centerline{\epsfxsize=3.0in\epsfbox{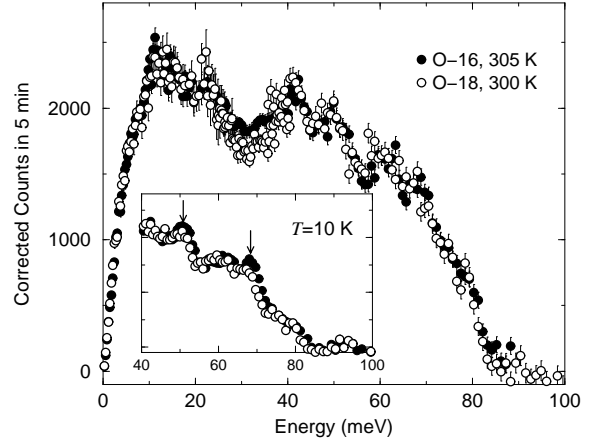}}
\caption[DOS for O-16 and O-18 at 300 K, inset 10 K]{Comparison of the
phonon density-of-states for $^{16}$O and $^{18}$O samples above $T_{C}$.
There are some slight differences which are more easily seen at low
temperature, shown in the inset.}
\end{figure}
$^{16}$O data
between 75 and 80~meV, where the signal at 305~K is slightly higher than at
225~K. A similar effect can be seen in the $^{18}$O data except that the
scatter in the data is larger. In other perovskites this energy region
corresponds to the oxygen breathing modes but there was no specific peak
seen in our data that could be identified with a mode of vibration in this
energy region. There may be a slight decrease in intensity of the peak near
51~meV in the $^{16}$O data. Both these trends are borne out by measurements
at 10~K in the 40 to 100 meV range. The published single crystal results %
\cite{ZhangPhonon} for the $x=0.3$ sample are far more conclusive and show
significant damping effects on the longitudinal Jahn-Teller (100) breathing
mode which has an energy of 72 meV at the zone center and 51 meV at the zone
boundary. Near the zone boundary the intensity of the phonons strongly
decreases with temperature until the mode disappears above $T_{C}$. This is
consistent with a decrease in the 51 meV peak seen in our data. Similar
changes are also seen in the Raman scattering.\cite{granado01} Overall,
however, the changes in the phonon density-of-states are modest.

A comparison between the different samples above $T_{C}$ also reveals that
they are nearly identical, as shown in Fig. 13 where we compare the phonon
density-of-states between samples rather than between temperatures. 
The different
temperatures were chosen to reflect the different $T_{C}$ values. The region
that looks most affected by oxygen isotope substitution at these
temperatures is between 20 and 40 meV where the scattering intensity near 30
meV is lower for the $^{18}$O sample than the $^{16}$O sample. Raman
measurements by Franck {\em et al.} \cite{franck98} near 0~K showed that the
oxygen modes at 28~meV and 53~meV are shifted downward by 1~meV and 2.6~meV
for the $^{18}$O sample. At low temperatures (inset) there does seem to be a
difference in the spectra at 68~meV where the $^{16}$O peak may be slightly
higher and more clearly defined. Overall, however, only small changes are
observed in the density-of-states, either with change in isotope, or with
change in temperature above versus below $T_{C}$, which is in agreement with
recent tunneling data.\cite{Hudspeth}

\section{Discussion}

The increase in the ferromagnetic transition when $^{18}$O is replaced by $%
^{16}$O, and the concomitant extension of the spin wave regime without any
change in the overall magnetic energetics, demonstrates in an unambiguous
way that the ferromagnetic state is prematurely truncated, and that the
transition from the ferromagnetic-metallic state to the
paramagnetic-insulator state is first order in nature. For the optimally
doped $x=3/8$ composition the extrapolated Curie point, where the transition
would be second order, is just above the observed ferromagnetic transition,
indicating that the transition is weakly first order at this composition. It
is this proximity to the second-order fixed point that enables the large ($%
\sim $30\%) decrease in the value of the spin stiffness upon $^{18}$O
replaced by $^{16}$O, making it unambiguous that the ferromagnetic state is
truncated. At somewhat higher Ca concentrations the transition in fact
becomes second order.\cite{Hellman}

The origin for the first-order truncation is the sudden formation of
polarons in the system, which trap the carriers and render the system
insulating. It is the polaron formation temperature that determines the
observed $T_{C}$ in the first-order regime, not the strength of the magnetic
interactions. Replacing $^{16}$O by $^{18}$O lowers the energy of the
Jahn-Teller phonon, making it easier to form polarons and consequently
lowers the transition temperature. The application of stress also lowers the
barrier for polaron formation. Lowering the Ca concentration below the $%
x=3/8 $ composition decreases the observed Curie temperature, while the
exchange energy is (initially) unaffected.\cite{fernandez98} This downturn
in $T_{C}({\rm observed})$ in the phase diagram then originates 
from lowering the
energy for polaron formation, as it becomes easier for the system to form
correlated Jahn-Teller distortions; recall that at $x=0$ pure LaMnO$_{3}$ is
strongly Jahn-Teller distorted, forming a structure with long range orbital
(and antiferromagnetic) order. The temperature range where the polaron glass
is observed \cite{Argyriouglass,Lynnglass} would then be expected to become
larger with decreasing $x$. This trend continues until the system becomes
insulating around $x\approx 0.2$, where the double-exchange interaction is
lost and the magnetic energetics drops precipitously.\cite%
{dai2001,fernandez2002}

Ultimately it is the structure that determines the energetics for polaron
formation, which can be cast in the form of the tolerance factor $t$.\cite%
{GoodenoughFO96,hwang95} The Ca system has a relatively small $t$ and forms
polarons relatively easily, and the polaron formation occurs in the
ferromagnetic state and causes a first-order transition. Larger ions such as
Sr or Ba have a substantially larger $t$, and do not appear to form a
significant number of polarons below $T_{C}$. The ferromagnetic transition
then becomes a conventional second-order transition, that is uncoupled from
the resistivity transition.\cite{KartikPRL,LynnBa} It will be interesting to
determine if there are dynamic polarons in the Sr and Ba substituted systems.

\section{Acknowledgments.}

We thank L. Vasiliu-Doloc for her assistance, and thank A. J. Millis, J. A.
Fernandez-Baca, P. Dai, and L. Vasiliu-Doloc for helpful discussions. Work
at the University of Maryland and Rutgers University was supported by 
the NSF-MRSEC Grant No. DMR
00-80008.  Work at NIST was supported in part by the Binational Science
Foundation, grant number 2000073.\
Work in Moscow was supported by ISTC
grant \#1859.















\begin{references}
\bibitem[*]{stfx} Present address: Department of Physics, St. Francis Xavier
University, Antigonish, N.S. B2G 2W5 Canada and email cadams@stfx.ca.

\bibitem[**]{towson} Present address: Department of Physics, Astronomy, and
Geosciences, Towson University, Towson, MD 21252-0001.

\bibitem{Wollan} E. O. Wollan and W. C. Koehler, Phys. Rev. {\bf 100, }545
(1955).

\bibitem{Goodenough} J. B. Goodenough, Phys. Rev. {\bf 100, }564 (1955).

\bibitem{tokurascience} Y. Tokura and N. Nagaosa, Science {\bf 288}, 462
(2000).

\bibitem{ramirez97} A. P. Ramirez, J. Phys.: Condens. Matter {\bf 9}, 8171
(1997).

\bibitem{lynn00} For a review of the spin dynamics see J. W. Lynn, J.
Supercond. \& Novel Mag. {\bf 13}, 263 (2000).

\bibitem{dagottoreview} E. Dagotto, T. Hotta, and A. Moreo, Phys. Rep. 
{\bf 344}, 1 (2001).

\bibitem{cheongreview} K. H. Kim, M. Uehara, V. Kiryukhin, and S.-W. Cheong,
cond-mat/0212113.

\bibitem{Theory} See, for example, A. J. Millis, P. B. Littlewood, and B. I.
Shraiman, Phys. Rev. Lett. {\bf 74}, 5144 (1995); H. R\"{o}der, J. Zang, and
A. R. Bishop, {\em ibid.} {\bf 76}, 1356 (1996); A. J. Millis, Phys. Rev. B 
{\bf 53}, 8434 (1996); C. M. Varma, {\em ibid.} {\bf 54}, 7328 (1996); A. S.
Alexandrov and A. M. Bratkovsky, Phys. Rev. Lett. {\bf 82}, 141 (1999); A.
Moreo, S. Yunoki, and E. Dagotto, Science {\bf 283}, 2034 (1999). P.
Schlottmann, Phys. Rev. B {\bf 62}, 439 (2000); D. I. Golosov,
Phys. Rev. Lett. {\bf 84}, 3974 (2000); J. van den Brink, G. Khaliullin, and
D. Khomskii, cond-mat/0206053; Y. Motome and N. Furukawa, cond-mat/0203041

\bibitem{Henniondrop} M. Hennion, F. Moussa, G. Biotteau, J.
Rodriguez-Carvajal, L. Pinsard, and A. Revcolevschi, Phys. Rev. Lett. {\bf 81%
}, 1957 (1998).

\bibitem{Granadodrop} E. Granado, C. D. Ling, J. J. Neumeier, J. W. Lynn,
and D. N. Argyriou (preprint).

\bibitem{zener51} C. Zener, Phys. Rev. {\bf 82}, 403 (1951).

\bibitem{zhao96} G. M. Zhao, K. Conder, H. Keller, and K. A. M\"{u}ller,
Nature (London) {\bf 381}, 676 (1996).

\bibitem{mori98} S. Mori, C. H. Chen, and S-W. Cheong, Nature (London) {\bf %
392}, 473 (1998); S. Mori, C. H. Chen, and S-W. Cheong, Phys. Rev. Lett. {\bf %
81}, 3972 (1998).

\bibitem{billinge96} S. J. L. Billinge, R. G. DiFrancesco, G. H. Kwei, J. J.
Neumeier, and J. D. Thompson, Phys. Rev. Lett. {\bf 77}, 715 (1996); D.
Louca, T. Egami, E. L. Brosha, H. R\"{o}der, and A. R. Bishop, Phys. Rev. B 
{\bf 56}, R8475 (1997).

\bibitem{alexandrov95} A. S. Alexandrov and N. F. Mott, {\em Polarons and
Bipolarons} (World Scientific, Singapore, 1995).

\bibitem{lynn96} J. W. Lynn, R. W. Erwin, J. A. Borchers, Q. Huang, A. Santoro,
J-L. Peng, and Z. Y. Li, Phys. Rev. Lett. {\bf 76}, 4046 (1996); J. W. Lynn,
R. W. Erwin, J. A. Borchers, A. Santoro, Q. Huang, J-L. Peng, and R. L.
Greene, J. Appl. Phys. {\bf 81}, 5488 (1997).

\bibitem{fernandez98} J. A. Fernandez-Baca, P. Dai, H. Y. Hwang, C. Kloc, and
S-W. Cheong, Phys. Rev. Lett. {\bf 80}, 4012 (1998).

\bibitem{adams2000} C. P. Adams, J. W. Lynn, Y. M. Mukovskii, A. A.
Arsenov, and D. A. Shulyatev, Phys. Rev. Lett. {\bf 85}, 3954 (2000).

\bibitem{dai2000} P. Dai, J. A. Fernandez-Baca, N. Wakabayashi, E. W. Plummer,
Y. Tomioka, and Y. Tokura, Phys. Rev. Lett. {\bf 85}, 2553 (2000).

\bibitem{doloc99} L. Vasiliu-Doloc, S. Rosenkranz, R. Osborn, S. K. Sinha,
J. W. Lynn, J. Mesot, O. H. Seeck, G. Preosti, A. J. Fedro, and J. 
F. Mitchell, Phys. Rev. Lett. {\bf 83}, 4393 (1999).

\bibitem{shimomura99} S. Shimomura, N. Wakabayashi, H. Kuwahara, and Y.
Tokura, Phys. Rev. Lett. {\bf 83}, 4389 (1999).

\bibitem{Kim2000} K. H. Kim, M. Uehara, and S-W. Cheong, Phys. Rev. B
{\bf 62}, R11945 (2000).

\bibitem{lynn01} J. W. Lynn, C. P. Adams, Y. M. Mukovskii, A. A. 
Arsenov, and D. A. Shulyatev, J. Appl. Phys. {\bf 89}, 6846 (2001).

\bibitem{Nelson} C. S. Nelson, M. v. Zimmermann, Y. J. Kim, J. P. Hill, D.
Gibbs, V. Kiryukhin, T. Y. Koo, S-W. Cheong, D. Casa, B. Keimer, Y. Tomioka,
Y. Tokura, T. Gog, and C. T. Venkataraman, Phys. Rev. B {\bf 64}, 174405
(2001).

\bibitem{Kiryukhin2002} V. Kiryukhin, T. Y. Koo, A. Borissov, Y. J. Kim, C.
S. Nelson, J. P. Hill, D. Gibbs, and S-W. Cheong, Phys. Rev. B {\bf 65},
094421 (2002).

\bibitem{Argyriouglass} D. N. Argyriou, J. W. Lynn, R. Osborn, B. Campbell,
J. F. Mitchell, U. Ruett, H. N. Bordallo, A. Wildes, and C. D. Ling, Phys.
Rev. Lett. {\bf 89}, 036401 (2002).

\bibitem{Lynnglass} J. W. Lynn, D. N. Argyriou, F. M. Woodward, Y. Ren, L.
Chapon, Y. M. Mukovskii, and D. A. Shulyatev (preprint).

\bibitem{schiffer95} P. Schiffer, A. P. Ramirez, W. Bao, and S.-W. Cheong,
Phys. Rev. Lett {\bf 75}, 3336 (1995).

\bibitem{cheongmax} S-W. Cheong and C. H. Chen, in {\em Colosssal
Magnetoresistance, Charge Ordering, and Related Properties of Manganese
Oxides}, edited by B. Raveau and C. N. R. Rao (World Scientific, New Jersey,
1998), p. 241.

\bibitem{smolyaninova01} V. N. Smolyaninova, A. Biswas, P. Fournier, S.
Lofland, X. Zhang, Guo-meng Zhao, and R. L. Greene, Phys. Rev. B {\bf 65},
104419 (2002).

\bibitem{HuangCacryst} Q. Huang, A. Santoro, J. W. Lynn, R. W. Erwin, J. A.
Borchers, J. L. Peng, K. Ghosh, and R. L. Greene, Phys. Rev. B
{\bf 58}, 2684 (1998).

\bibitem{Hennion2001} G. Biotteau, M. Hennion, F. Moussa, J.
Rodriguez-Carvajal, L. Pinsard, A. Revcolevschi, Y. M. Mukovskii, and D.
Shulyatev, Phys. Rev. B {\bf 64}, 104421 (2001).

\bibitem{lynn95} For a review of the experimental technique see J. W. Lynn
and J. A. Fernandez-Baca, in {\em The Magnetism of Amorphous Metals and Alloys%
}, edited by J. A. Fernandez-Baca and W-Y. Ching (World Scientific,
Singapore, 1995), Chap. 5, p. 221.

\bibitem{doloc98} L. Vasiliu-Doloc, J. W. Lynn, A. H. Moudden, A. M. de
Leon-Guevara, and A. Revcolevschi, Phys. Rev. B {\bf 58}, 14913 (1998).

\bibitem{privman91} V. Privman, P. C. Hohenberg, and A. Aharony, in {\em %
Phase Transitions and Critical Phenomena}, edited by C. Domb and J. L.
Lebowitz (Academic, New York, 1991), p. 1

\bibitem{okuda2000} T. Okuda, Y. Tomioka, A. Asamitsu, and Y. Tokura,
Phys. Rev. B {\bf61}, 8009 (2000).

\bibitem{dai2001} P. Dai, J. A. Fernandez-Baca, E. W. Plummer, Y. Tomioka,
and Y. Tokura, Phys. Rev. B {\bf 64}, 224429 (2001).

\bibitem{heffner2times} R. H. Heffner, J. E. Sonier, D. E. MacLaughlin, G.
J. Nieuwenhuys, G. Ehlers, F. Mezei, S.-W. Cheong, J. S. Gardner, and H.
R\"{o}der, Phys. Rev. Lett. {\bf 85}, 3285 (2000).

\bibitem{deteresa97} J. M. deTeresa, M. R. Ibarra, P. A. Algarabel,
C. Ritter,
C. Marquina, J. Blasco, J. Garcia, A. del Moral, and Z. Arnold, Nature
(London) {\bf 386}, 256 (1997).

\bibitem{GoodenoughFO96} W. Archibald, J.-S. Zhou, and J. B. Goodenough,
Phys. Rev. B {\bf 53}, 14445 (1996).

\bibitem{SANSref} J. S. Higgins and H. Benoit, {\em Polymers and Neutron
Scattering}, Oxford (1994).

\bibitem{pyrochlore} See, for example, J. W. Lynn, L. Vasiliu-Doloc, and M.
A. Subramanian, Phys. Rev. Lett. {\bf 80}, 4582 (1998).

\bibitem{Hellman} D. Kim, B. Revaz, B. L. Zink, F. Hellman, J. J. Rhyne, and
J. F. Mitchell, Phys. Rev. Lett. {\bf 89}, 227202 (2002).

\bibitem{CMRrefs} R. von Helmolt, J. Wecker, B. Holzapfel, L. Schultz, and
K. Samwer, Phys. Rev. Lett. {\bf 71}, 2331 (1993); S. Jin, T. H. Tiefel, M.
McCormack, R. A. Fastnacht, R. Ramesh, and L. H. Chen, Science {\bf 264},
413 (1994).

\bibitem{2dstrain} A. Biswas, M. Rajeswari, R. C. Srivastava, Y. H. Li, T.
Venkatesan, R. L. Greene, and A. J. Millis, Phys. Rev. B {\bf 61}, 9665
(2000); A. Biswas, M. Rajeswari, R. C. Srivastava, T. Venkatesan, 
R. L. Greene,
Q. Lu, A. L. de Lozanne, and A. J. Millis, {\em ibid.} {\bf 63}, 184424
(2001); J. O'Donnell, M. S. Rzchowski, J. N. Eckstein, and I. Bozovic, Appl.
Phys. Lett. {\bf 72}, 1775 (1998); X. W. Wu, M. S. Rzchowski, H. S. Wang,
and Qi Li, Phys. Rev. B {\bf 61}, 501 (2000).

\bibitem{hydro} J. J. Neumeier, M. F. Hundley, J. D. Thompson, and R. H.
Heffner, Phys. Rev. B {\bf 52}, R7006 (1995).

\bibitem{hwang95} H. Y. Hwang, S-W. Cheong, P. G. Radaelli, M. Marezio, and
B. Batlogg, Phys. Rev. Lett. {\bf 75}, 914 (1995).

\bibitem{bandwidthCMR}
H. Kuwahara, Y. Moritomo, Y. Tomioka, A. Asamitsu, 
M. Kasai, R. Kumai, and Y. Tokura, Phys. Rev. B {\bf 56}, 9386 (1997); 
H. Kuwahara, Y. Tomioka, Y. Moritomo, A. Asamitsu, 
M. Kasai, R. Kumai, and Y. Tokura, Science {\bf 272}, 80 (1996); 
Y. Tokura, H. Kuwahara, Y. Moritomo,
Y. Tomioka, and A. Asamitsu, Phys. Rev. Lett. {\bf 76}, 3184 (1996).

\bibitem{isaac98} I. Isaac and J. P. Franck, Phys. Rev. B {\bf 57}, R5602
(1998).

\bibitem{Babushkina} N. A. Babushkina, L. M. Belova, O. Yu. Gorbenko, A. R.
Kaul, A. A. Bosak, V. I. Ozhogin, and K. I. Kugel, Nature (London) 
{\bf 391}, 159 (1998).

\bibitem{franck98} J. P. Franck, I. Isaac, Weimin Chen, J. Chrzanowski, and
J. C. Irwin, Phys. Rev. B {\bf 58}, 5189 (1998); J. P. Franck, I. Isaac,
Weimin Chen, J. Chrzanowski, J. C. Irwin, and C. C. Holmes, J. 
Supercond. {\bf 12}, 263 (1999).

\bibitem{ZhouGood} J.-S. Zhou and J. B. Goodenough, Phys. Rev. Lett. {\bf 80}%
, 2665 (1998).

\bibitem{Heilman} A. K. Heilman, Y. Y. Xue, Y. Y. Sun, R. L. Meng, Y. S.
Wang, B. Lorenz, C. W. Chu, J. P. Franck, and W. Chen, Phys. Rev. B {\bf 61},
8950 (2000).

\bibitem{Babushkina2000} N. A. Babushkina, A. N. Taldenkov, L. M. Belova, E.
A. Chistotina, O. Y. Gorbenko, A. R. Kaul, K. I. Kugel, and D. I. Khomskii,
Phys. Rev. B {\bf 62}, 6081 (2000).

\bibitem{Fisher} R. A. Fisher, F. Bouquet, N. E. Phillips, J. P. Franck, G.
Zhang, J. E. Gordon, and C. Marcenat, Phys. Rev. B {\bf 64}, 134425 (2001).

\bibitem{rhyne} J. J. Rhyne, H. Kaiser, H. Luo, G. Xiao, and M. L. Gardel,
J. Appl. Phys. {\bf 83, }7339{\bf \ }(1998); J. J. Rhyne, H. Kaiser, L.
Stumpe, J. F. Mitchell, T. McCloskey, and A. R. Chourasia, {\em ibid.} 
{\bf 87}, 5813 (2000); L. Stumpe, B. Kiry, H. Kaiser, J. J. Rhyne, and J. F.
Mitchell, {\em ibid.} {\bf 91}, 7511 (2002).

\bibitem{fernandez2002} J. A. Fernandez-Baca, P. Dai, H. Kawano-Furukawa, H.
Yoshizawa, E. W. Plummer, S. Katano, Y. Tomioka, and Y. Tokura, Phys. Rev. B
{\bf 66}, 054434 (2002).

\bibitem{millis96} A. J. Millis, B. I. Shraiman, and R. Mueller, Phys. Rev.
Lett. {\bf 77}, 175 (1996).

\bibitem{Sumarlin} I. W. Sumarlin, J. W. Lynn, D. A. Neumann, J. J. Rush,
C-K. Loong, J. L. Peng, and Z. Y. Li, Phys. Rev. B {\bf 48}, 473 (1993).

\bibitem{ZhangPhonon} J. Zhang, P. Dai, J. A. Fernandez-Baca, E. W. Plummer,
Y. Tomioka, and Y. Tokura, Phys. Rev. Lett. {\bf 86}, 3823 (2001).

\bibitem{granado01} E. Granado, A. Garci{\' a}, J. A. Sanjurjo, C. Rettori,
and I. Torriani, Phys. Rev. B {\bf 63}, 64404 (2001).

\bibitem{reichardt99} W. Reichardt and M. Braden, Physica B {\bf 263}, 416
(1999).

\bibitem{dai2000prb} P. Dai, 
H. Y. Hwang, Jiandi Zhang, J. A. Fernandez-Baca, S.-W. Cheong, 
C. Kloc, Y. Tomioka, and Y. Tokura
Phys. Rev. B {\bf 61}, 9553 (2000).

\bibitem{Hudspeth} H. D. Hudspeth, F. Sharifi, I. J. Guilaran, P. Xiong, and
S. von Molnar, Phys. Rev. B {\bf 65}, 052405 (2002).

\bibitem{KartikPRL} K. Ghosh, C. J. Lobb, R. L. Greene, S. G. Karabashev, D.
A. Shulyatev, A. A. Arsenov, and Y. Mukovskii, Phys. Rev. Lett. {\bf 81},
4740 (1998).

\bibitem{LynnBa} J. W. Lynn, L. Vasiliu-Doloc, K. Ghosh, S. Skanthakumar, S.
N. Barilo, G. L. Bychkov, and L. A. Kurnevitch (preprint).

\end{references}
\end{document}